\documentclass[sigconf,natbib=false]{acmart}

\usepackage{booktabs} 
\usepackage{microtype} 
\usepackage{enumitem}
\usepackage{siunitx}
\usepackage{xspace} 
\usepackage{soul} 
\usepackage{adjustbox} 
\usepackage{subcaption}
\usepackage{tabularx}
\usepackage{array,multirow}
\usepackage{colortbl}
\usepackage{threeparttable} 
\usepackage{tablefootnote}
\usepackage{comment}
\usepackage{tikz} 
\usepackage{wrapfig}
\usetikzlibrary{patterns}
\usetikzlibrary{shapes,arrows}
\usetikzlibrary{positioning,calc}
\usetikzlibrary{decorations.pathreplacing}
\usetikzlibrary{shapes.geometric,shapes.arrows,decorations.pathmorphing}
\usetikzlibrary{matrix,chains,scopes,positioning,arrows,fit,backgrounds}
\usepackage{longtable}
\usepackage[utf8]{inputenc}
\usepackage{tcolorbox}
\usepackage[mode=buildnew]{standalone}
\usepackage{soul}
\usepackage{listings} 
\usepackage{mathtools}
\usepackage{amsmath}
\usepackage{graphicx} 

\usepackage[ruled]{algorithm2e}

\usepackage{makecell}
\usepackage{pifont}

\usepackage[version=4]{mhchem} 

\definecolor{codegreen}{rgb}{0,0.6,0}
\definecolor{codegray}{rgb}{0.5,0.5,0.5}
\definecolor{codepurple}{rgb}{0.58,0,0.82}
\definecolor{backcolour}{rgb}{0.95,0.95,0.92}

\lstdefinestyle{mystyle}{
  backgroundcolor=\color{backcolour}, commentstyle=\color{codegreen},
  keywordstyle=\color{magenta},
  numberstyle=\tiny\color{codegray},
  stringstyle=\color{codepurple},
  basicstyle=\ttfamily\footnotesize,
  breakatwhitespace=false,         
  breaklines=true,                 
  captionpos=b,                    
  keepspaces=true,                 
  numbers=left,                    
  numbersep=5pt,                  
  showspaces=false,                
  showstringspaces=false,
  showtabs=false,                  
  tabsize=2
}
\usepackage{xcolor}
\usepackage{mdframed}
\usepackage{xcolor}
\usepackage[strict]{changepage}

\usepackage{framed}

\definecolor{formalshade}{rgb}{0.824, 0.973, 0.824}

\newenvironment{formal}{%
  \MakeFramed{\advance\hsize-\width\FrameRestore}%
  \noindent\hspace{-4.55pt}
  \begin{adjustwidth}{}{3pt}%
}
{%
  \vspace{1pt}\end{adjustwidth}\endMakeFramed%
}

\AtBeginDocument{%
  }

\setcopyright{acmlicensed}

\RequirePackage[
  datamodel=acmdatamodel,
  style=acmnumeric,
  ]{biblatex}

\begin{document}


\title{Sustainable Quantum Computing: Opportunities and Challenges of Benchmarking Carbon in the Quantum Computing Lifecycle}


\author{Nivedita Arora}
\email{nivedita@northwestern.edu}
\affiliation{%
  \institution{Northwestern University}
  \country{USA}
}

\author{Prem Kumar}
\email{kumarp@northwestern.edu}
\affiliation{%
  \institution{Northwestern University}
  \country{USA}
}

\renewcommand{\shortauthors}{Arora et al.}

\begin{abstract}
 While researchers in both industry and academia are racing to build Quantum Computing (QC) platforms with viable performance and functionality, the environmental impacts of this endeavor, such as its carbon footprint, e-waste generation, mineral use, and water and energy consumption, remain largely unknown. A similar oversight occurred during the semiconductor revolution and continues to have disastrous consequences for the health of our planet. As we build the quantum computing stack from the ground up, it is crucial to comprehensively assess it through an environmental sustainability lens for its entire life-cycle: production, use, and disposal (\autoref{fig:full_life_QC}). In this paper, we highlight the need and challenges in establishing a QC sustainability benchmark that enables researchers to make informed architectural design decisions and celebrate the potential `quantum environmental advantage.' We propose a Carbon-aware Quantum Computing (CQC) framework that provides the foundational methodology and open research questions in calculating the total life-cycle carbon footprint of a QC platform. Our call to action to the research community is the establishment of a new research direction known as ``sustainable quantum computing" that promotes both quantum computing for sustainability-oriented applications and the sustainability of quantum computing. 
\end{abstract}

%
\vspace{-0.1in}


\received{15 December 2023}

\maketitle

\section{Introduction } 
\begin{figure}[!ht]
\vspace{0in}
\centering
\includegraphics[height=2.5in]{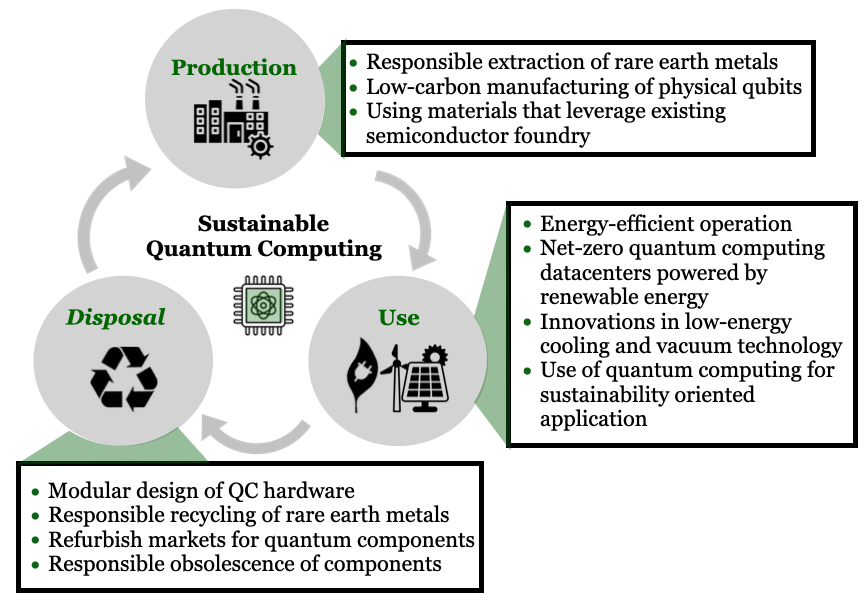}
\vspace{-0.2in}
\caption{Sustainable Quantum Computing: \textmd{The need for proactive benchmarking of the environmental effects of quantum computing across its entire lifecycle: production, use, and disposal is critical.}}
\vspace{-0.2in}
\label{fig:full_life_QC}

\end{figure}

\begin{formal}
\centering In the qubit revolution, sustainability needs to be a forethought, not an afterthought.
\end{formal}

Quantum computing (QC) holds the potential to solve computational problems that lie beyond the reach of classical semiconductor computers but are crucial to the societal advancement, e.g., simulation of new materials, drug discovery, supply chain optimization, and cryptography \cite{feynman2018simulating, preskill2018quantum}. Currently, quantum computers require significant energy expenditures to maintain low-temperature and vacuum-pressure operational environments \cite{ladd2010quantum}, constantly need error-correcting operations, and utilize ecologically sensitive resources such as rare-earth metals and noble gases. Despite these resource-intensive factors with significant ecological impact, the progress of quantum computing is currently only benchmarked in terms of speed and functionality. The dire consequences of a similar neglect of the environment during the semiconductor revolution are evident today: electronic waste is the fastest growing waste stream globally \cite{forti_global_2020, osibanjo_challenge_2007}. The Information and Communication Technology (ICT) sector contributed to 11\% of global energy consumption in 2020, and is projected to increase to 21\% by 2030 driven by the growing demand for cloud computing \cite{jones2018stop}. In 2021, US data centers' water consumption was equivalent to that of five million US households over the year  \cite{awwa_water_use, water_footprint_calculator}. We cannot afford to repeat such oversight during the quantum revolution. Even though today's quantum computing landscape resembles the maturity of the traditional mainframe semiconductor technology from 1960s (e.g., room-size computers and nascent efforts of building internet) \cite{roberts1988arpanet}, we still advocate for a proactive approach towards environmental sustainability, and not post-hoc remedial steps. 


\subsection*{Contributions}
\begin{enumerate}[leftmargin=0.25cm]
    \item \textbf{Necessity and challenges of benchmarking sustainability in QC platforms:} To promote responsible development of quantum technology, quantifying its environmental impact is a crucial first step. We delve into the advantages of sustainability benchmarking in QC and the complexities of undertaking this task (\autoref{sec:need_challenges}). 

    \item \textbf{Proposed Carbon-aware Quantum Computing (CQC) framework}: Understanding the ecological footprint of quantum computing throughout its entire lifecycle (production, use, i.e., operation and application, and end-of-life recycle/disposal) is essential for minimizing resource use and environmental impact. Uniquely, QC also holds the potential to significantly reduce environmental burdens in other industries, such as optimizing fertilizer production. To encompass this holistic view, we create Carbon-aware Quantum Computing (CQC) framework that proposes to quantify the carbon footprint using carbon equivalence for each lifecycle phase including application. We also discuss opportunities and open research questions for carbon footprint reduction at each stage (\autoref{sec:cqc}).
    
 \item \textbf{Call to action}: Our call to action is the establishment of "sustainable quantum computing" as a research sub-field, fostering research teams encompassing diverse disciplines to advance both the sustainability of quantum computing and its applications in sustainability (\autoref{sec:call}).

\end{enumerate}
\begin{formal}
Quantum computing needs a \textbf{sustainability initiative} that benchmarks its carbon emissions throughout its entire life cycle (production, use, and disposal) and informs computing stack design decisions.
\end{formal}

We hope that this paper sparks the interest of a broad set of research communities that would like to start thinking about sustainable quantum computing (SQC). To bridge the knowledge gap, we provide a primer on relevant Quantum Computing (\autoref{sec:qc}) and Life Cycle Analysis (\autoref{sec:lca}) fundamentals, in addition to our contributions.

\section{Primer on Quantum Computing}\label{sec:qc}
The scientific realization that nature does not behave classically but quantum mechanically gave rise to a new field of research in the early 1980s known as Quantum Computing (QC) \cite{ benioff1980computer,  feynman2018simulating}. 
\begin{figure}[!ht]
\centering

\includegraphics[height=1.2in]{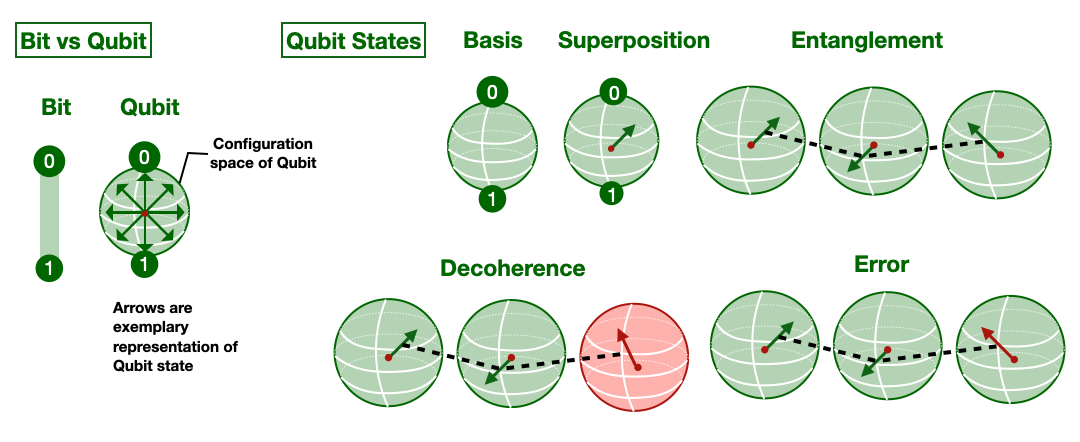}
\vspace{-0.15in}
\caption{ \textmd{Difference between a bit and a qubit and different qubit states.}}
\label{fig:qubitstates}
\vspace{-0.15in}
\end{figure}

\subsubsection*{Qubit States (\autoref{fig:qubitstates}):}
Unlike classical computer bits that are confined to the 0 or 1 state, quantum computers utilize qubits that defy the binary world. They are not just 0 or 1, but a blur of possibilities, enabling it to be both 0 and 1 simultaneously. Imagine a coin `spinning so fast' that it has both heads and tails, i.e., \textit{superposition}. Each qubit dances within a \textit{basis}, a reference frame defining its head and tail (like spin up/down). A qubit collapses from superposition to a basis state when measured. Qubits can also be \textit{entangled}, where the state of one qubit becomes directly related to the state of another, regardless of the distance between them. These unique principles of superposition and entanglement bolster the high information density and the \textit{quantum promise}--- the potential to solve a certain type of problems much more efficiently than current semiconductor-based classical computers. Environmental interactions and imperfections in the quantum system can cause a qubit in superposition or entanglement to collapse, a process called \textit{decoherence}, which limits the computation time. To elongate this time, the qubits need to be operated in extremely controlled environments and constantly error corrected. Even with that, \textit{errors} can still arise from imperfections in the hardware, noise in the control signals, or limitations in our ability to manipulate the qubits.

\begin{figure}[!ht]
\centering

\includegraphics[height=3.5in]{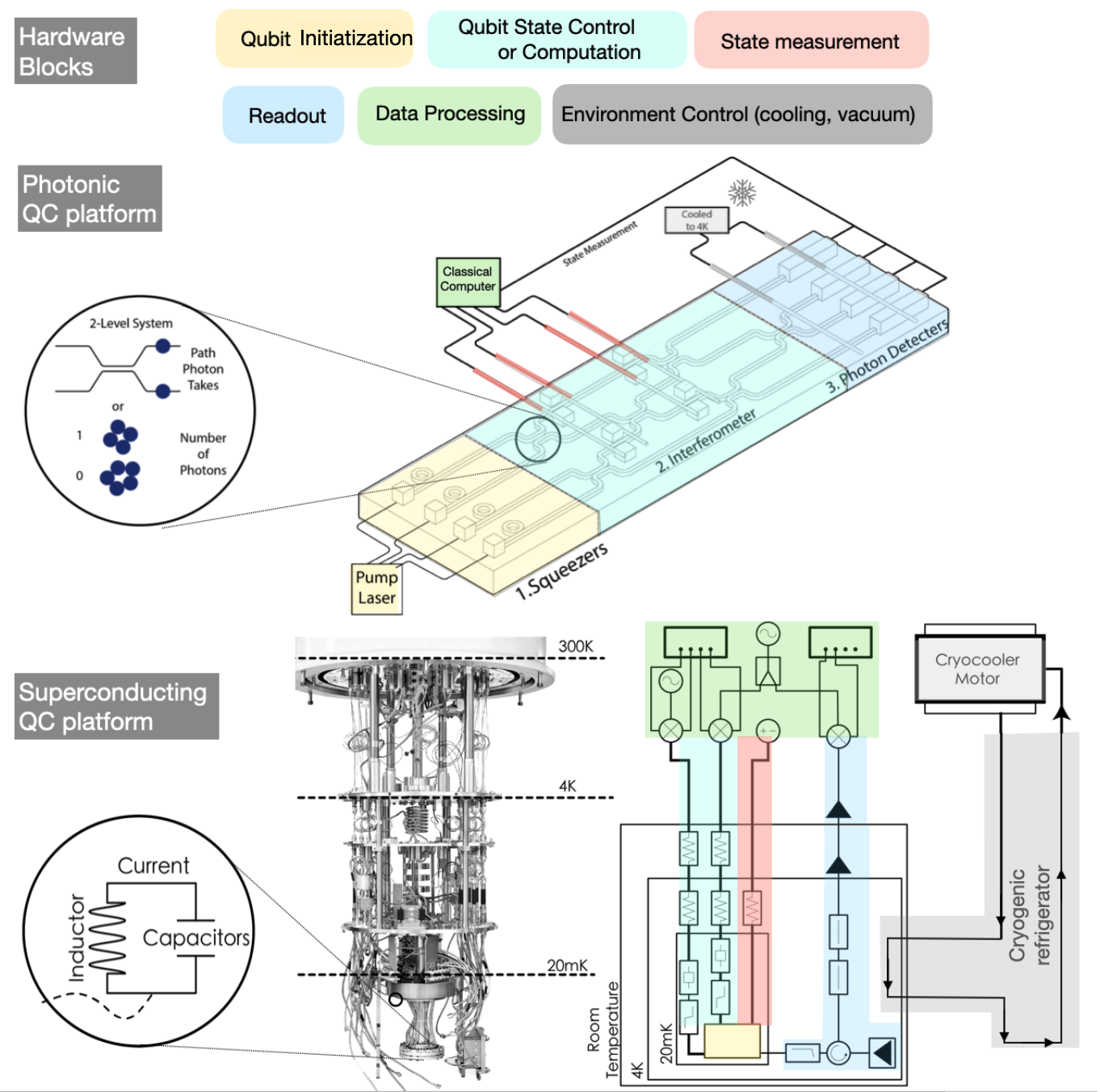}
\vspace{-0.1in}
\caption{ \textmd{Key QC hardware blocks and their illustration in photonic and superconducting QC platforms.}}
\label{fig:qc_hardware}

\end{figure}

\subsubsection*{QC hardware architecture}
A physical qubit can be generated by controlling atoms, photons, or electrons \cite{li2018survey}. Some of the most mature QC platforms that are in production are trapped-ions, neutral atoms, superconducting, and photonics. 
 \autoref{fig:qc_hardware} illustrates the key component blocks of quantum hardware architecture, using photonic and superconducting QC platforms as examples. \textit{Qubit hardware} is responsible for generating and housing the initialized physical qubits. For instance, in photonic QC, laser pulses driving a squeezer generate/initialize the qubits. 
 In superconducting systems, resonator loops with Josephson junctions fulfill this role. The \textit{qubit state control} hardware performs the actual computation and error correction. At its core, a quantum algorithm is a series of operations that manipulate the state of qubits from one configuration to another. During algorithm execution, the qubit states are continuously \textit{manipulated} and error-corrected. Finally, the qubit states are measured, i.e., \textit{read out}, and the data is sent to a classical computer or super-computer for \textit{processing}. Depending on the qubit architecture, different hardware blocks require \textit{cryogenic cooling and/or vacuum-pressure} conditions. For example, in a photonics platform the readout needs 4K operational temperature, while in a superconducting platform, qubit hardware is at 20mK, while qubit control, measurement, and readout are at 4K. 
 \vspace{-0.1in}   
\subsubsection*{Progress of QC platforms} Current QC platforms, called Noisy Intermediate Scale Quantum (NISQ), are still rudimentary with too few qubits and high error probabilities \cite{preskill2018quantum}. Quantum error correction leading to Fault-Tolerant Quantum Computing (FTQC) \cite{fruchtman2016technical}, architectures needed to unlock the quantum promise as they function, are still far away from realization.

\section{Necessity and Challenges of QC sustainability benchmarking}\label{sec:need_challenges}
The expanding landscape of quantum computing platforms, each with its own resource demands, qubit scale, and quality, demands a nuanced approach to platform selection for a computational task. For example, if QC platform 1 performs a computational task 2x faster than QC platform 2 but also consumes 10x more energy and 5x more water for operation, would we still call it a win for humanity to choose platform 1? If QC platform 3 uses 10,000x more energy in its manufacturing than platform 4 but platform 3 has the capability to solve much more computationally complex tasks, then which one should be chosen? Being able to quantify the environmental impact of a QC platform across production, use, and disposal with application complexity and benefit in mind is the path forward.  Developing such environmental benchmarks for QC is a complex undertaking due to the following key challenges: 

\begin{enumerate}[leftmargin=0.25cm]
    \item \textbf{Fragmentation in QC platforms:} With a diverse set of QC platforms, each with its unique hardware architecture and tailored quantum algorithm implementation, it is hard to understand resource demands. For example, one platform's cryogenic cooling requirements do not match the other; one platform may need just a laptop for data processing, while the other may need a supercomputer.  
    \item \textbf{Lack of universal performance benchmarks}: Unlike the well-established FLOPS metric used for a classical computer's performance,  effective performance benchmarks for QC are an area of active research \cite{eisert2020quantum}. This makes creating metrics like Watts/FLOP or $MT_{CO_{2e}}$/FLOP hard. Instead, we need to rely on different QC benchmark tasks, e.g., gate fidelity, quantum Fourier transform. 
    \item \textbf{Bringing manufacturing, use, disposal, and application benefits under a common metric:} Individually quantifying these different parameters is one task, but we can only fairly compare when they all are brought under the same comparison metric. 
\end{enumerate}


\section{Primer on Life Cycle Analysis}\label{sec:lca}
To tackle challenge 3 mentioned in \autoref{sec:need_challenges}, we will ground our QC environmental benchmarking framework (\autoref{sec:cqc}) on an established field of research called Life Cycle Analysis (LCA) \cite{guinee2002handbook}. LCA is a methodology for the systematic and quantitative evaluation of the environmental performance of a product through all stages of its life cycle. LCA uses `Equivalent-carbon or $CO_{2e}$' as its quantitative measure. Life Cycle Inventory (LCI) database \cite{nistlci2020} quantifies $CO_{2e}$ of different environmental resources (e.g., wood, water, coal), greenhouse gases (e.g., methane, nitrogen oxide, He-3), and processed materials (e.g., steel, cement). LCA discusses the total carbon footprint of a product in two forms -- 
\begin{itemize}[leftmargin=0.25cm]
    \item Operational Carbon: $CO_{2e}$ of resources like energy, water, and materials used in the operation of the product. Thus, energy from renewable sources is preferred more than from coal.   
    \item Embodied Carbon: This is $CO_{2e}$ of product other than use, such as extraction of minerals, manufacturing, transport, repair, disposal.
\end{itemize}

With the stress on sustainability of the past 5-10 years \cite{whitehouse2021cop26}, different ICT technologies---datacenters \cite{acun2023carbon, gupta2022act}, artificial intelligence \cite{wu2022sustainable}, internet of things \cite{arora2022circularity, prakash2023tinyml}, laptops and AR/VR \cite{apple2022macbookproedp}---have been quantified for their operational and embodied carbon, and their trade-offs with performance and functionality. 

\section{Carbon-aware Quantum Computing}\label{sec:cqc}
Inspired by the LCA of semiconductor-based ICT, we propose the Carbon-aware Quantum Computing (CQC) framework. In the equation below, there are three parts of the total carbon. The embodied and operational carbon is the same as that of LCA. They promote the `sustainability for QC' platforms. The last term is for application-centric carbon, which inspires `QC for sustainability.'

\vspace{15pt}
\begin{mdframed}[backgroundcolor=formalshade] 
        CQC: \hspace{3pt} $Total_{CO_{2e}}$= $Embodied_{CO_{2e}}$ + $Operational_{CO_{2e}}$ \\ \hspace*{80pt}- $Application_{CO_{2e}}$ \hspace*{30pt}
\end{mdframed}
\vspace{10pt}
Next, we discuss the potential sources and steps to quantify and optimize operational (\autoref{5.1} and \autoref{5.2}), embodied (\autoref{5.3} and \autoref{5.4}), and application-centric (\autoref{5.5}) carbon in a QC stack. 

\subsection{Operational Energy}\label{5.1}
Determining the energy consumption of a specific computational task on a particular QC platform with a specific qubit scale, quality, and speed, is an open research question with little exploration. Recently, a thermodynamical model of a universal QC has demonstrated an exponential energy advantage over classical supercomputers for Simon's problem \cite{meier2023energy}. Such simulations often fail to consider the energy expenditures of real physical hardware control and data processing  \cite{jaschke2023quantum}. Vice versa, just understanding hardware components fails to capture algorithmic nuances \cite{elsayed2019review}. Balancing the both, Fellous-Asiani {\em et al.} showed how a superconducting computer could reach desired accuracy within a given energy budget \cite{fellous2023optimizing}. There is a lot more work to be done to understand the energy consumption of different QC platforms \cite{auffeves2022quantum}, and their energy vs. performance tradeoffs. For this, we suggest, first solving challenge 1 (\autoref{sec:need_challenges}) by creating a systematic catalog of the operational power consumed by different QC hardware blocks (\autoref{sec:qc}) and their components. This data can be sourced from product datasheets, experimental results reported in research papers, and, where necessary, theoretical physics models. The next step would be dividing the computational algorithm as a series of qubit state transitions (\autoref{sec:qc}). 
Using the transition times between two qubit states, which are well documented in the literature as gate operation durations, we can calculate the energy. Finally, repeating this process for all possible state transitions allows us to estimate the system-level energy required for a specific computational task (Joule). Furthermore, we can extend this analysis to compute the average power consumption per qubit for that task (Watts/qubit).


\subsection{Operational Energy to Operational Carbon}\label{5.2} 
While operational energy efficiency is an important metric, it does not guarantee operational carbon efficiency \cite{shenoy2023energy}. This opens a series of research questions related to the source of energy and water utilized in quantum computers (QCs) and its effect on the operational carbon footprint. 
\begin{itemize}[leftmargin=0.25cm]
\item \textbf{Carbon-neutral operation of quantum computers:} In the United States, renewable energy generation is projected to increase from 20\% in 2020 to 42\% by 2050 as the nation pursues Net Zero \cite{eo14076}. QC should also moon-shot towards carbon-neutral operation. It is important to think, which type of QC platform would likely achieve carbon-neutral operation first and for which type of computational operations? It is important to be strategic about the selection of locations of future QC datacenters, in order to maximumize renewable energy operation. 
\item \textbf{Optimisation of QC architecture to handle energy spikes:} Energy harvested from renewable sources like sunlight or wind is inherently variable, fluctuating in availability throughout the day and season. How can we optimize the hardware and software architecture of QCs to adapt to and efficiently utilize these clean energy sources?
\end{itemize}


\vspace{-0.1in}
\subsection{Embodied Carbon for Production}\label{5.3} 

\begin{figure}[!ht]
\centering
\includegraphics[height=2in]{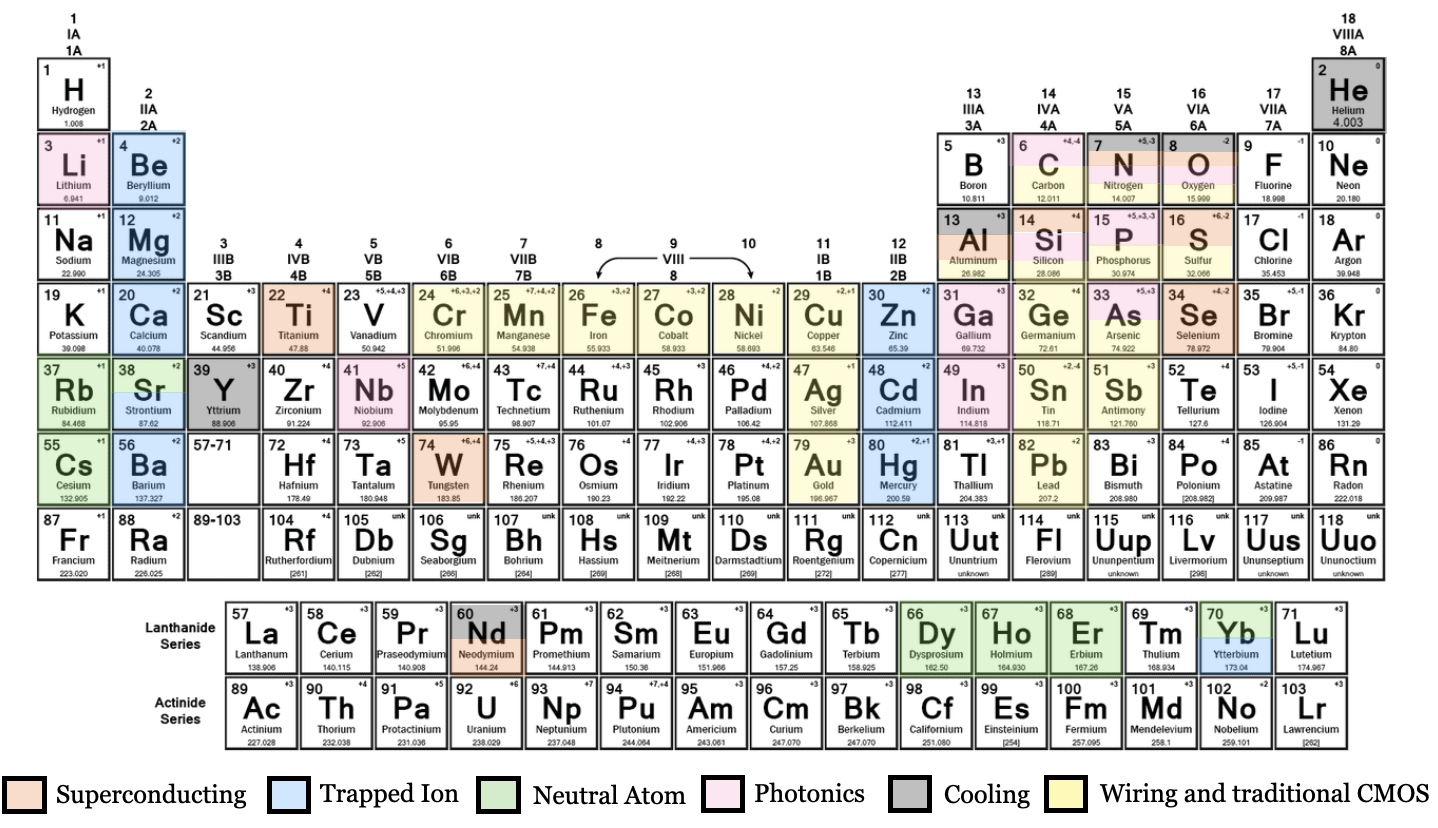}

\caption{\textmd{Materials and minerals used in 1) physical qubits,  2) cryogenic cooling, 3) qubit controls and data processing.}}
\label{fig:periodic_table}
\end{figure}
\vspace{-0.15in}
Several rare earth metals are crucial for developing quantum platforms, but they also come with a significant environmental cost (\autoref{fig:periodic_table}). For example, dysprosium (Dy) plays a vital role in neutral atom qubits, and Helium-3 is essential for cryogenic cooling \cite{de2021materials}. Comparing them to greener alternatives will illuminate the sustainability vs. performance trade-offs, guiding future development.


\subsection{Embodied Carbon for Disposal}\label{5.4} 
Graceful degradation of technology is as important as production. As certain components in quantum computers (QCs) age, they can become less efficient and consume significantly more energy. Replacing such components becomes environmentally advantageous at a certain point \cite{acun2023carbon}. Assessing carbon friendly age of component obsolescence is a critical tradeoff between embodied and operational carbon. Building quantum computing (QC) hardware in a modular fashion would significantly reduce resource requirements for repair. Salvaging reusable components from QC hardware for other industries could extend their lifespan and benefit diverse fields without adding additional carbon. Responsible extraction and reuse of rare earth elements from obsolete QC hardware is another way to reduce carbon footprint.

\subsection{Application-centric Carbon Offset}\label{5.5} 

\begin{figure}[!ht]
\centering
\includegraphics[height=0.65in]{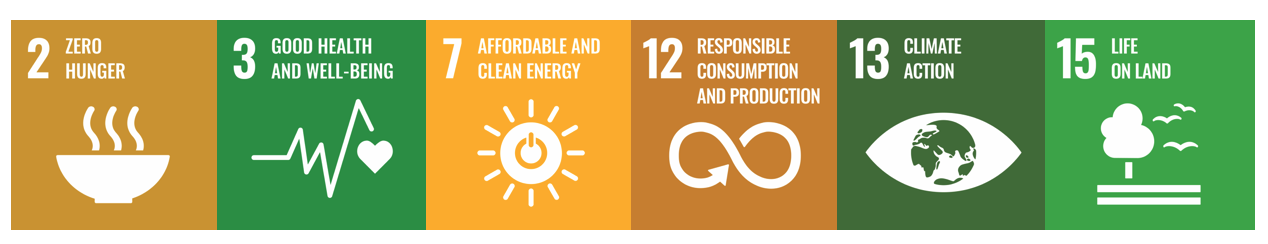}
\caption{\textmd{QC has the potential to support many of the UN SDGs and create huge negative carbon footprint offset.}}
\label{fig:unsdg}
\end{figure}

Paris Agreement \cite{ParisAgreement} and SDGs \cite{un2015sdgs} have driven technology's role in sustainability. Quantum computing is uniquely positioned to solve complex basic science problems that could lead to a huge reduction in global energy consumption and carbon footprint. For example, it can improve 
the efficiency of the Haber process used in the manufacturing of fertilizers (\#2), help in drugs discovery (\#3), model climate change (\#13), simulate battery chemistry and nuclear fusion reaction for cleaner energy (\#7) and optimize the supply chain (\#15). $Application_{CO_{2e}}$ adjusts these in the total carbon footprint of QC. This promotes the development of QC platforms 
even if they have high initial embodied and operational carbon. Calculating these carbon offsets requires QC and LCA researchers to work together with domain experts.

\section{Community Call to action }\label{sec:call}

\begin{figure}[!ht]
\centering
\includegraphics[height=0.45in]{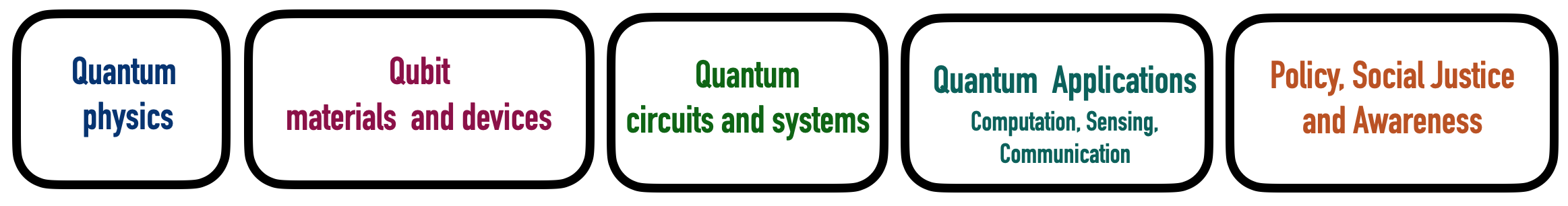}
\vspace{-0.1in}
\caption{\textmd{Need for building a cross-stack interdisciplinary community working towards sustainable quantum computing.} }
\label{fig:community}

\end{figure}
We believe that setting up the sustainable quantum computing research agenda requires a cross-stack interdisciplinary community working together (\autoref{fig:community}).
\begin{itemize}[leftmargin=0.25cm]
\item {\textbf{Researchers:}} Though QC platforms are in their infancy, integrating environmental responsibility throughout development and application is crucial. Collaboration between QC and sustainability researchers is key to bolstering sustainable QC. 

\item \textbf{Educators:} Championing sustainability alongside performance is crucial for QC's future. Sustainability-focused QC education will empower the next generation of engineers to prioritize environmental impact, not just performance.

\item \textbf{Industry:} Just like traditional ICT Environmental Product Reports \cite{apple2022macbookproedp}, upcoming commercial QC platforms should apply the CQC benchmark to generate a carbon footprint report. 

\item \textbf{International agencies and policymakers:} QC, like other semiconductor-based ICT, relies on import-export of rare earths. International agencies prioritize QC \cite{nqiact2018, qtpartners2023, ostp2023qc}, but sustainable mining and usage for QC must be added to align with UNSDG 17's global partnership for sustainable development.

\end{itemize}

\section{Conclusion}
This paper champions a vision for quantum computing that transcends the singular pursuit of fault tolerance by prioritizing the creation of a carbon-neutral, fault-tolerant quantum computer that solves sustainability-oriented applications. For this, we advocate for proactively benchmarking the environmental effects of the QC platform for its entire lifecycle: production, use (operation, application), and disposal and leveraging the information to guide architectural decisions. We introduce the Carbon-aware Quantum Computing (CQC) framework for calculating the total carbon footprint of a quantum computer and highlight the opportunities for carbon reduction at each stage. Finally, we discuss the need for academia, industry, and funding agencies to collaborate in establishing a sustainability initiative for quantum computing.


\printbibliography

@article{preskill2018quantum,
  title={Quantum computing in the NISQ era and beyond},
  author={Preskill, John},
  journal={Quantum},
  volume={2},
  pages={79},
  year={2018},
  publisher={Verein zur F{\"o}rderung des Open Access Publizierens in den Quantenwissenschaften}
}

@article{feynman2018simulating,
  title={Simulating physics with computers},
  author={Feynman, Richard P and others},
  journal={Int. j. Theor. phys},
  volume={21},
  number={6/7},
  year={2018}
}

@incollection{roberts1988arpanet,
  title={The Arpanet and computer networks},
  author={Roberts, Larry},
  booktitle={A history of personal workstations},
  pages={141--172},
  year={1988}
}

@article{ladd2010quantum,
  title={Quantum computers},
  author={Ladd, Thaddeus D and Jelezko, Fedor and Laflamme, Raymond and Nakamura, Yasunobu and Monroe, Christopher and O’Brien, Jeremy Lloyd},
  journal={nature},
  volume={464},
  number={7285},
  pages={45--53},
  year={2010},
  publisher={Nature Publishing Group UK London}
}

@misc{water_footprint_calculator,
  author= {Water Footprint Calculator},
  url = {https://www.watercalculator.org/},
   year ={2023},
}

@misc{awwa_water_use,
  author = {American Water Works Association},
  title = {Average Daily {US} Residential Water Use},
  url = {https://www.awwa.org/Portals/0/AWWA%20Resources/Water%20Education/Average-Daily-Residential-Water-Use.pdf},
  year ={2023}
}

@article{benioff1980computer,
  title={The computer as a physical system: A microscopic quantum mechanical Hamiltonian model of computers as represented by Turing machines},
  author={Benioff, Paul},
  journal={Journal of statistical physics},
  volume={22},
  pages={563--591},
  year={1980},
  publisher={Springer}
}

@article{fruchtman2016technical,
  title={Technical roadmap for fault-tolerant quantum computing},
  author={Fruchtman, Amir and Choi, Iris},
  journal={NQIT Technical Roadmap},
  year={2016}
}

@online{ParisAgreement,
title = {{Paris Agreement}},
year = {2015},
url = {{https://unfccc.int/process-and-meetings/what-is-the-united-nations-framework-convention-on-climate-change}},
organization = {{United Nations Framework Convention on Climate Change}},
urldate = {2023-12-13},

}

@misc{eo14076,
  title = {Executive Order 14076: Tackling the Climate Crisis at Home and Abroad},
  Author = {The White House},
  date = {2021-01-27},
  url = {https://www.whitehouse.gov/briefing-room/presidential-actions/2021/01/27/executive-order-on-tackling-the-climate-crisis-at-home-and-abroad/},
  urlaccess = {2023-12-15},
}

@online{un2015sdgs,
  title = {Transforming our world: the 2030 Agenda for Sustainable Development},
  publisher = {United Nations},
  date = {2015-09-25},
  url = {https://sdgs.un.org/2030-agenda},
  urlaccess = {2023-12-15},
}

@legal{nqiact2018,
  Author = {The {US} Congress},
  title = {National Quantum Initiative Act of 2018},
  number = {Pub. L. No. 115-312},
  date = {2018-12-21},
  url = {https://www.congress.gov/bill/115th-congress/house-bill/6227/text},
  urlaccess = {2023-12-15},
}

@online{qtpartners2023,
  title = {Quantum Technology Partnership (QTP)},
  publisher = {Quantum Technology Partnership},
  date = {2023-12-15},
  url = {https://qt-partners.com/},
  urlaccess = {2023-12-15},
}

@article{shenoy2023energy,
  title={Energy-Efficiency versus Carbon-Efficiency: What's the difference?},
  author={Shenoy, Prashant},
  journal={ACM SIGENERGY Energy Informatics Review},
  volume={2},
  number={4},
  pages={1--2},
  year={2023},
  publisher={ACM New York, NY, USA}
}

@article{li2018survey,
  title = {A Survey on quantum computing technology},
  author = {Li, Yan and Gong, Su-Peng and Yin, Zhi-Guo and Lu, Zhi-Yuan and Zhang, Wen-Qi and Gong, Wei-Min and Xu, Fei and Lin, Guang-Can and You, Zhi and Shi, Feng-Xia and Li, Jian-Wei and Peng, Chang and Zhang, Xiao-Min and Shi, Yan and Wu, Wei and Guo, Guang-Can and Wang, Chao and Guo, Guo-Ping and Chen, Zhi-Chao and Pan, Jian-Wei},
  journal = {Frontiers of Physics},
  volume = {13},
  number = {11},
  pages = {114001},
  year = {2018},
  url = {https://doi.org/10.1007/s11467-018-09404-y},
  urlaccess = {2023-12-15},
}

@article{de2021materials,
  title={Materials challenges and opportunities for quantum computing hardware},
  author={De Leon, Nathalie P and Itoh, Kohei M and Kim, Dohun and Mehta, Karan K and Northup, Tracy E and Paik, Hanhee and Palmer, BS and Samarth, Nitin and Sangtawesin, Sorawis and Steuerman, David W},
  journal={Science},
  volume={372},
  number={6539},
  pages={eabb2823},
  year={2021},
  publisher={American Association for the Advancement of Science}
}

@report{ostp2023qc,
  title = {Quantum Computing for American Prosperity},
  Author = {The White House Office of Science and Technology Policy},
  date = {2023-05-04},
  url = {https://www.whitehouse.gov/briefing-room/statements-releases/2022/05/04/fact-sheet-president-biden-announces-two-presidential-directives-advancing-quantum-technologies/},
  urlaccess = {2023-12-15},
}

@article{jones2018stop,
  title={How to stop data centres from gobbling up the world’s electricity},
  author={Jones, Nicola and others},
  journal={Nature},
  volume={561},
  number={7722},
  pages={163--166},
  year={2018},
  publisher={Springer Science and Business Media LLC}
}

@techreport{forti_global_2020,
	title = {The {Global} {E}-waste {Monitor} 2020: {Quantities}, flows, and the circular economy potential},
	url = {http://ewastemonitor.info/download-2020/},
	institution = {United Nations University/United Nations Institute for Training and Research},
	author = {Forti, Vanessa and Balde, Cornelis Peter and Kuehr, Ruediger and Bel, Garam},
	year = {2020},
	pages = {120},
}

@article{osibanjo_challenge_2007,
	title = {The challenge of electronic waste (e-waste) management in developing countries},
	volume = {25},
	issn = {0734-242X},
	url = {https://doi.org/10.1177/0734242X07082028},
	doi = {10.1177/0734242X07082028},
	language = {en},
	number = {6},
	urldate = {2021-11-14},
	journal = {Waste Management \& Research},
	author = {Osibanjo, O. and Nnorom, I.C.},
	month = dec,
	year = {2007},
	pages = {489--501},
}

@database{nistlci2020,
  title = {NIST Building and Construction Life Cycle Inventory Database (GaBi)},
  publisher = {National Institute of Standards and Technology (NIST)},
  year = {2020},
  url = {https://www.nist.gov/el/sustainabledesign/lifecycle-assessment/lci-databases},
  note = {Version 7.1 accessed December 15, 2023},
}

@online{whitehouse2021cop26,
title = {FACT SHEET: President Biden Renews {U.S.} Leadership on World Stage at U.N. Climate Conference (COP26)},
Author = {The White House},
date = {2021-11-01},
url = {https://www.whitehouse.gov/briefing-room/statements-releases/2021/11/01/fact-sheet-president-biden-renews-u-s-leadership-on-world-stage-at-u-n-climate-conference-cop26/},
urlaccess = {2023-12-15},
}

@article{wu2022sustainable,
  title={Sustainable ai: Environmental implications, challenges and opportunities},
  author={Wu, Carole-Jean and Raghavendra, Ramya and Gupta, Udit and Acun, Bilge and Ardalani, Newsha and Maeng, Kiwan and Chang, Gloria and Aga, Fiona and Huang, Jinshi and Bai, Charles and others},
  journal={Proceedings of Machine Learning and Systems},
  volume={4},
  pages={795--813},
  year={2022}
}

@online{apple2022macbookproedp,
  title = {Environmental Product Declaration: 13-inch MacBook Pro},
  Author = {Apple Inc.},
  url = {https://www.apple.com/environment/pdf/products/notebooks/13-inch_MacBook_Pro_PER_June2022.pdf},
  date = {2022},
  format = {pdf},
}

@article{prakash2023tinyml,
  title={Is TinyML Sustainable?},
  author={Prakash, Shvetank and Stewart, Matthew and Banbury, Colby and Mazumder, Mark and Warden, Pete and Plancher, Brian and Reddi, Vijay Janapa},
  journal={Communications of the ACM},
  volume={66},
  number={11},
  pages={68--77},
  year={2023},
  publisher={ACM New York, NY, USA}
}

@inproceedings{arora2022circularity,
  title={Circularity in Energy Harvesting Computational "Things"},
  author={Arora, Nivedita and Iyer, Vikram and Oh, Hyunjoo and Abowd, Gregory D and Hester, Josiah D},
  booktitle={Proceedings of the 20th ACM Conference on Embedded Networked Sensor Systems},
  pages={931--933},
  year={2022}
}

@inproceedings{acun2023carbon,
  title={Carbon explorer: A holistic framework for designing carbon aware datacenters},
  author={Acun, Bilge and Lee, Benjamin and Kazhamiaka, Fiodar and Maeng, Kiwan and Gupta, Udit and Chakkaravarthy, Manoj and Brooks, David and Wu, Carole-Jean},
  booktitle={Proceedings of the 28th ACM International Conference on Architectural Support for Programming Languages and Operating Systems, Volume 2},
  pages={118--132},
  year={2023}
}

@inproceedings{gupta2022act,
  title={ACT: Designing sustainable computer systems with an architectural carbon modeling tool},
  author={Gupta, Udit and Elgamal, Mariam and Hills, Gage and Wei, Gu-Yeon and Lee, Hsien-Hsin S and Brooks, David and Wu, Carole-Jean},
  booktitle={Proceedings of the 49th Annual International Symposium on Computer Architecture},
  pages={784--799},
  year={2022}
}

@article{eisert2020quantum,
  title={Quantum certification and benchmarking},
  author={Eisert, Jens and Hangleiter, Dominik and Walk, Nathan and Roth, Ingo and Markham, Damian and Parekh, Rhea and Chabaud, Ulysse and Kashefi, Elham},
  journal={Nature Reviews Physics},
  volume={2},
  number={7},
  pages={382--390},
  year={2020},
  publisher={Nature Publishing Group UK London}
}

@book{guinee2002handbook,
  title={Handbook on life cycle assessment: operational guide to the ISO standards},
  author={Guin{\'e}e, Jeroen B},
  volume={7},
  year={2002},
  publisher={Springer Science \& Business Media}
}

@article{auffeves2022quantum,
  title={Quantum technologies need a quantum energy initiative},
  author={Auffeves, Alexia},
  journal={PRX Quantum},
  volume={3},
  number={2},
  pages={020101},
  year={2022},
  publisher={APS}
}

@article{fellous2023optimizing,
  title={Optimizing resource efficiencies for scalable full-stack quantum computers},
  author={Fellous-Asiani, Marco and Chai, Jing Hao and Thonnart, Yvain and Ng, Hui Khoon and Whitney, Robert S and Auff{\`e}ves, Alexia},
  journal={PRX Quantum},
  volume={4},
  number={4},
  pages={040319},
  year={2023},
  publisher={APS}
}

@article{meier2023energy,
  title={Energy-Consumption Advantage of Quantum Computation},
  author={Meier, Florian and Yamasaki, Hayata},
  journal={arXiv preprint arXiv:2305.11212},
  year={2023}
}

@inproceedings{elsayed2019review,
  title={A review of quantum computer energy efficiency},
  author={Elsayed, Nelly and Maida, Anthony S and Bayoumi, Magdy},
  booktitle={2019 IEEE Green Technologies Conference},
  pages={1--3},
  year={2019},
  organization={IEEE}
}

@article{jaschke2023quantum,
  title={Is quantum computing green? An estimate for an energy-efficiency quantum advantage},
  author={Jaschke, Daniel and Montangero, Simone},
  journal={Quantum Science and Technology},
  volume={8},
  number={2},
  pages={025001},
  year={2023},
  publisher={IOP Publishing}
}


\end{document}